\documentclass[twocolumn,showpacs,showkeys,amsmath,amssymb]{revtex4}

\usepackage{graphicx}

\begin{document}


\title{Stock price fluctuations and the mimetic behaviors of traders}

\author{Jun-ichi Maskawa}
 \affiliation{Department of Management Information, Fukuyama Heisei University, Fukuyama, Hiroshima 720-0001,
 Japan}
 \email{maskawa@heisei-u.ac.jp}

\date{\today}

\begin{abstract}
We give a stochastic microscopic modelling of stock markets driven
by continuous double auction. If we take into account the mimetic
behavior of traders, when they place limit order, our virtual
markets shows the power-law tail of the distribution of returns
with the exponent outside the Levy stable region, the short memory
of returns and the long memory of volatilities. The Hurst exponent
of our model is asymptotically 1/2. An explanation is also given
for the profile of the autocorrelation function, which is
responsible for the value of the Hurst exponent.
\end{abstract}


\maketitle

\section{Introduction}
In financial markets, it seems very natural to believe that large
price changes are caused by large transaction volumes
\cite{gabaix2003}. Farmer et al. has, however, proposed an
entirely different story of large price changes in the markets
driven by continuous double auctions \cite{farmer2004}. They has
argued that large returns are not caused by large orders, while
the large gaps between the occupied price levels in the orderbook
lead to large price changes in each transaction, and actually
showed that the gap distribution closely matches the return
distribution, based on the analysis of the orderbook as well as
the transaction records on the London Stock Exchange. They have
also shown that the virtual market orders of a constant size
reproduces the actual distribution of returns.

Then, we arrive at the next question, that is, what really causes
large gaps. Maslov has introduced and studied a simple model of
markets driven by continuous double auctions \cite{maslov2000}. In
his model, traders choose one from the two types of orders at
random. One is a market order which is a order to sell or buy a
fixed amount of shares immediately at the best available price on
the market on the time. The other is a limit order which is a
order to buy or sell a fixed amount of shares with the
specification of the limit price (the worst allowable price). The
relative price of a new limit order to the most recent market
price is a stochastic variable drawn from a uniform distribution
in a given interval. Numerical simulations of the model show that
it has such realistic features to a certain extent as the
power-law tail of the distribution of returns, and the long range
correlation of the volatility. However the price evolution has
some essentially different statistical properties from the actual
one on the market of large investors. First the exponent $\alpha$
of the power-law tail of the distribution of returns is inside the
Levy stable region ( $0 < \alpha \leq 2$ ), while the actual value
is close to 3 \cite{gopik1998,plerou1999}. Second the Hurst
exponent H = 1/4 is unrealistic in wide enough time windows. In
actual market, we have H = 1/2 for long-term, which is the value
for free diffusion. Challet and Stinchcombe have proposed a model
with non-constant occurrence rates of various types of orders as a
model in the same class as Maslov's \cite{challet2003}. The price
of their model is over diffusive ($H>1/2$) for short-term and
diffusive (H=1/2) for long-term.

In this paper, we propose a stochastic model  with a novel feature
in the same class as the above models. We take into account the
mimetic behavior of traders, when they place limit order, then our
virtual markets shows the behaviors which are widely recognized as
the stylized facts in financial markets, that is, the power-law
tail of the distribution of returns with the exponent outside the
Levy stable region \cite{gopik1998,plerou1999}, the short memory
of returns and the long memory of volatilities \cite{liu1999}. The
Hurst exponent of our model is asymptotically 1/2. An explanation
is also given for the profile of the autocorrelation function,
which is responsible for the value of the Hurst exponent.

\section{Model}
We give the definition of our model here. We introduce the
cancellation of orders to maintain the number of orders stored in
orderbook as in the model of Smith et al \cite{smith2003}. and in
other models \cite{challet2003}. We have, therefore, three types
of orders in total on both sides of trade, namely sell/buy market
order, sell/buy limit order and the cancellation of sell/buy limit
order. The conservation law indicates $\alpha_i -\mu_i-\delta_i=0$
for i = sell or buy, where the parameters $\alpha_i$, $\mu_i$ and
$\delta_i$ denote the occurrence rate of limit order, market order
and cancellation respectively. For simplicity, we also assume that
traders always place a fixed size of order.

The prices at which new limit orders are placed reflect traders'
strategies. In Maslov's paper \cite{maslov2000}, they are
determined by offsetting the most recent market price by a random
amount drawn from a uniform distribution in a given interval.
Challet and Stinchcombe assume the Gaussian distribution from
which random variables for the relative prices of sell/buy limit
order to ask/bid are drawn, and put the exclusion hypothesis that
if there is already a order at the price, the price is not
available \cite{challet2003}. In the model of Smith et al.
\cite{smith2003}, traders place the sell(buy) limit order at any
price in the semi-infinite range above bid (below ask) with
uniform probability.

In this paper, we take a little bit more strategic order placement
than previous works. We assume the mimetic behavior of traders,
when they place limit orders. Traders in our model sometimes (or a
part of traders always) leave the decision of the limit price to
the others. They believe the majority, and should be so patient
that they can wait in line the execution of their orders for long
time. The assumption is implemented here as follows: with
probability $p$, a limit price of sell(buy) order is chosen with
the probability proportional to the size of orders stored on the
price, and with probability 1-$p$, a price is randomly chosen
between [bid+1, ask+1] ([bid-1, ask-1]) with uniform probability.
Parameter $p$ is crucial for the model. The choice of the limit
price in our model follows the preferential attachment dynamics in
the growth model of scale free networks
\cite{barabasi1999,albert2002}. ``Rich gets richer'' is a concept
common to both models. However, our model is not a growth model,
but the total amount of limit orders is loosely fixed, because the
number of coming limit orders are balanced with market orders and
cancellations. Instead, the result of simulations will show that
the distribution of the fluctuation of the gaps between occupied
price levels has a power-law tail owing to the unequal attractive
powers of each of the prices.

\section{Numerical simulations of the model}

The models of continuous double auctions such as Maslov's and ours
have a difficulty to solve analytically due to the existence of
free boundaries: sell (buy) limit prices acceptable for traders
who place the orders are bounded on the best price of the opposite
side, namely the bid (ask) price on the time. Slanina
\cite{slanina2001} and Smith et al. \cite{smith2003} have
formulated a mean field approximation of the master equations to
have good results. We adopt, however, a numerical method, and
leave the analytical examination for the coming papers.

We generate the each types of orders in the ratio of
$\alpha_i=0.25$ (limit order), $\mu_i=0.125$ (market order) and
$\delta_i=0.125$ (cancellation) for $i$ either equal to sell or
buy. For the several values of $p$, we perform 1,000 times runs of
10,000 step iterations with different initial conditions. We place
a unit of shares at each price 1, 2, -1, -2 as the ask, the second
best sell limit price, the bid and the second best buy limit price
respectively, and also 198 units of sell (buy) limit orders at
random in the range between 1 (-1) and the random integer drawn
from a uniform distribution in the interval [2,201] ([-201,-2]) as
a initial condition.

Here we present the most interesting results obtained through the
numerical simulations. Fig. 1 shows the cumulative distribution
functions of price shifts, the gaps between ask and the second
best sell limit price and spreads. Price shifts and the gaps are
sampled after every buy market order. The results for sell market
order are omitted here in order to avoid a redundancy, because we
take the symmetric values of the parameters and the initial
conditions. Spreads are sampled after every sell and buy market
orders. All the three distributions become broader when the
parameter $p$ becomes larger. The power law tails appear in all
the graphs when $p$ beyond 0.4. The distributions for the
parameter $p$ beyond 0.5 are very broad, but steep falls are
observed in the tail. We pick up some points from the interval
$0.4 < p < 0.5$, and roughly estimate that the power law exponent
of the tails have minima in the interval $0.45 < p < 0.5$, and the
values are near 3 at $p$=0.475 as given in the caption of Fig. 1.
We use the Hill estimator of the largest $\sqrt{n}$ data, where n
is the size of sample.

We see from Fig. 2 that the relative limit price, namely the
distance at which new limit orders are placed away from the
current best price, is broadly distributed when $p$ becomes large.
We want to demonstrate, however, that the broadness itself of the
distribution of the relative limit price does not create the fat
tail of the price shift distribution. First of all, for the
purpose, we collect the data of the limit order by a numerical
simulation of the model. Then we shuffle the order of arrivals of
limit orders, and perform a simulation using the surrogate data
instead of the limit price generated by the original rule of our
model. The comparison of the resultant probability distribution of
price shift of the surrogate data with that of the original data
is shown in Fig. 3. The tail of the distribution does not show a
power law behavior, though the original data does. This experiment
reveals that the information of orderbook plays a essential role
in the decision of the price at which new orders are placed in our
model. A similar role of the orderbook will be expected even in
real markets, though the style of the reference to the orderbook
is possibly different from that assumed here.

\section{Autocorrelation functions of the model}

We derive the autocorrelation functions of price shift and of the
absolute value of the price shift obtained by the numerical
simulations of the model. The results are given in the panels of
Fig. 4. including comparison with the autocorrelation functions of
the surrogate data mentioned in the previous section. In those
panels, the unit of time increment corresponds to a buy market
order.

The autocorrelation function of price shift almost vanishes except
the value of time lag $\tau=1$ for both data. The values of the
autocorrelation at time lag $\tau=1$ are -0.41 and -0.46
respectively. Those values are close to -0.5, and are explainable
by the mean field approximation of the autocorrelation function as
follows: let $\delta_1$ and $\delta_2$ denote the mean square root
of price shift normalized by the standard deviation. The value
$\delta_1$ measures the price shift across the spread,
corresponding to the case that the side of the trade changes from
bid to ask or from ask to bid. The value $\delta_2$ corresponds to
the case that the side of the trade remains the same. If we assume
that the four cases occurs with the same probability 1/4 , the
mean field approximation of autocorrelation functions gives the
equation $\rho_i=<dp_t
dp_{t+i}>/\sigma^2=1/4(\delta_2^2-\delta_1^2)\delta_{1i}$. From
the normalization condition $\delta_1^2+\delta_2^2=2$ and the
inequality $\delta_1>>\delta_2$ (because spread always exists,
while the trade successively occurred on the same side do not
necessarily move the price), we have the result
$\rho_i\approx-0.5\delta_{1i}$. The profile of the autocorrelation
function is responsible for the value of the Hurst exponent H
through the equation $Var(p_t-p_0)/\sigma^2=Var(\sum_{i=1}^t
dp_i)/\sigma^2=t+\sum_{i=1}^t (t-i)\rho_i\sim t^{2H}$. In such
case of short memory as our model, we have the equation
$Var(p_t-p_0)/\sigma^2=Dt\sim t^{2H}$ for $t>>1$. In our case, the
diffusion constant $D=1+2\sum_{i=1}^t \rho_i$ is quite small owing
to the equation $\rho_i\approx-0.5\delta_{1i}$, and H=1/2 for
large t. An empirical study of the price diffusion is presented in
Fig. 5.

We see from the panel (b) of Fig. 4 that the autocorrelation
functions of the absolute value of price shift (empirical
volatility) have long memory. Both data plotted there are well
fitted by power laws. The original data, however, hold the memory
of volatility stronger than the surrogate data does.

\section{Conclusions}
Taking the strategy leaving the decision of the limit price to the
others in the stochastic model of financial markets driven by
continuous double auction, the virtual market shows the power-law
tail of the distribution of returns with the exponent near 3
according to the parameter which determines the ratio of the
mimetic limit order. The short memory of returns and the long
memory of volatilities are also reproduced by the model. The Hurst
exponent H of our model is asymptotically 1/2. The mean field
approximation explains the profile of the autocorrelation
function, which is responsible for the value of the Hurst exponent
H. The strategy assumed here are effective in holding the memory
of market volatility strong.

The author thanks D. Challet for attracting my notice to their
papers. He learns a lot from them.


\newpage

\begin{figure*}
\includegraphics[width=6cm]{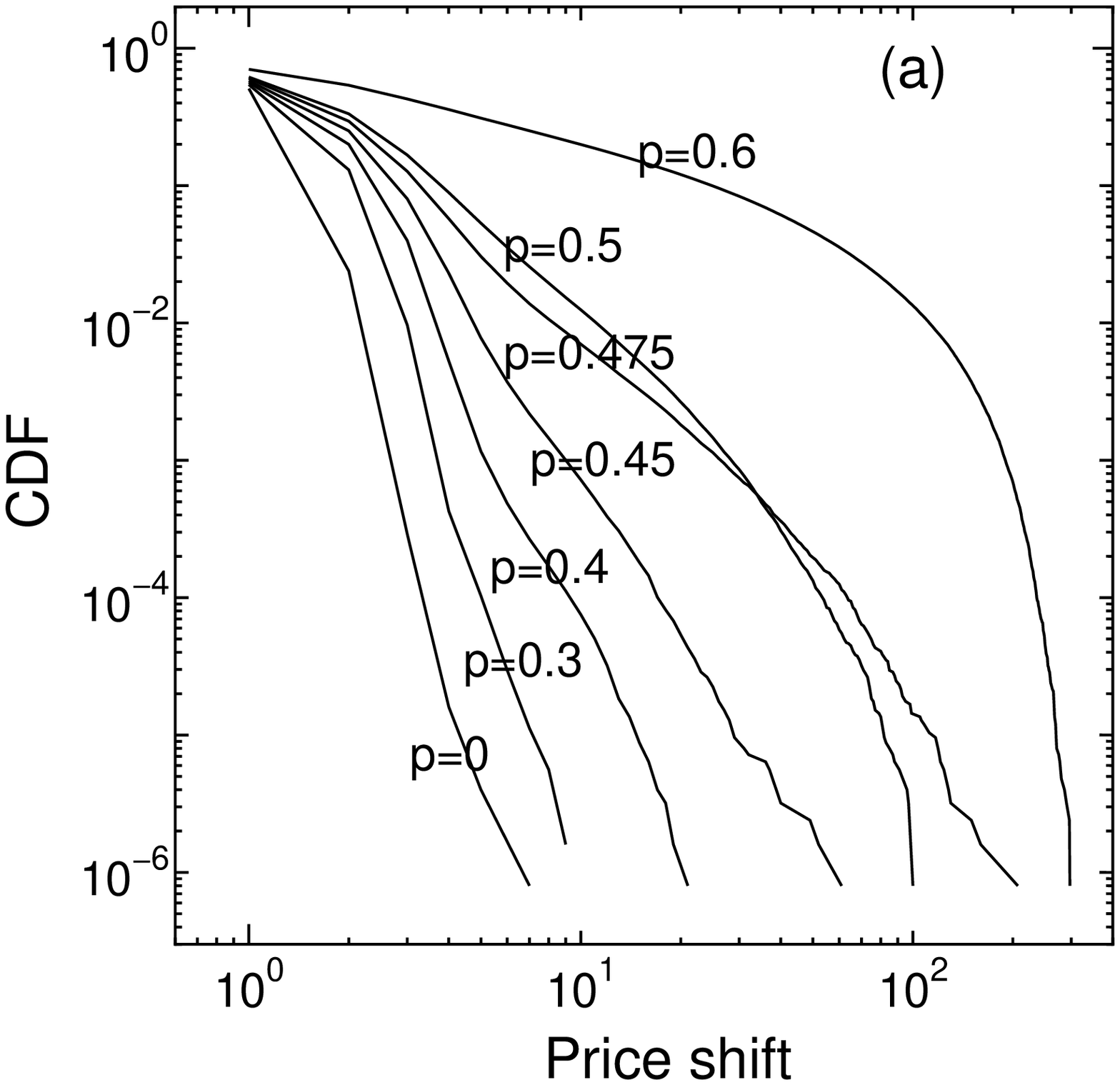}
\end{figure*}
\begin{figure*}
\includegraphics[width=6cm]{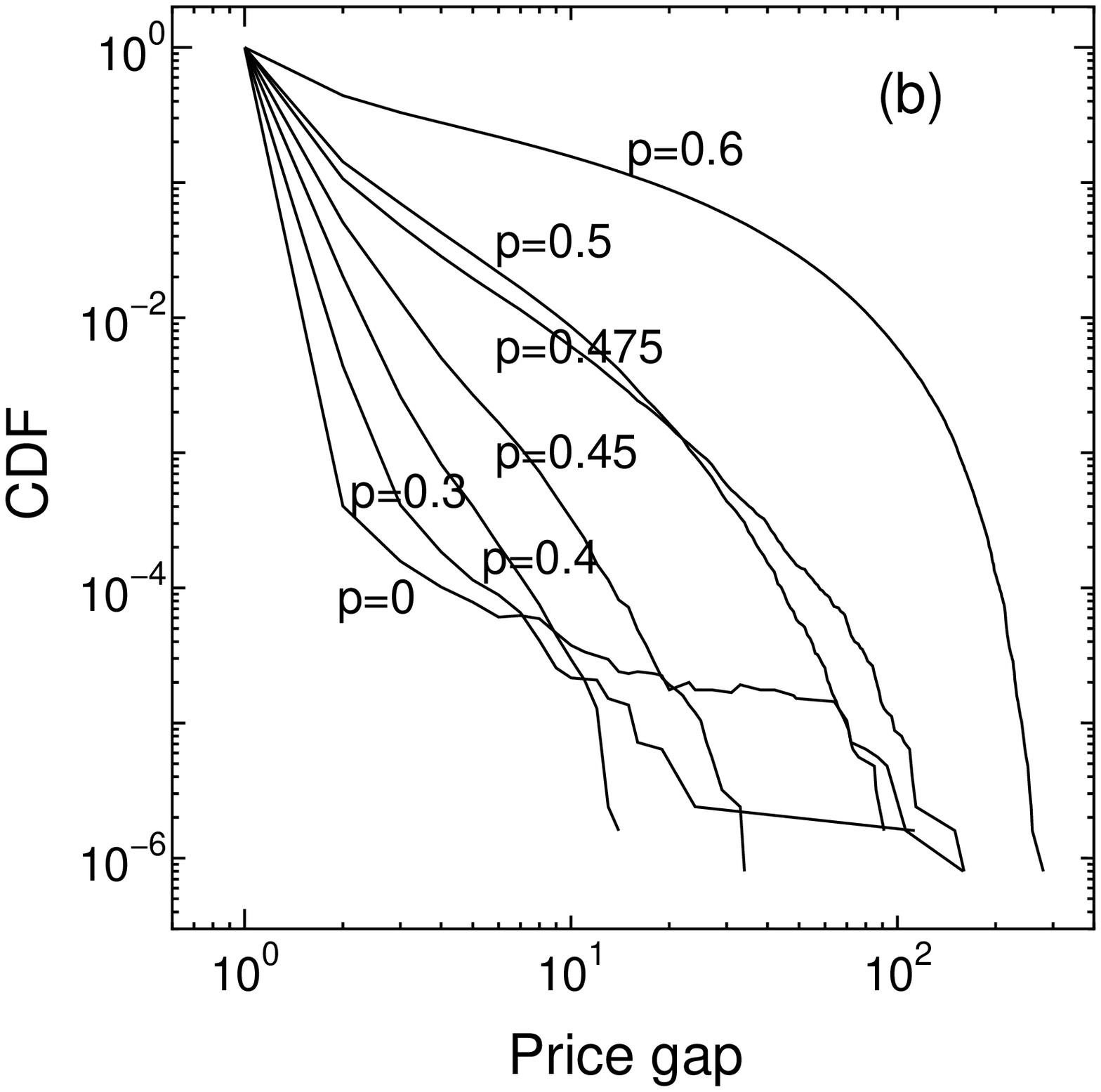}
\end{figure*}

\begin{figure*}
\includegraphics[width=6cm]{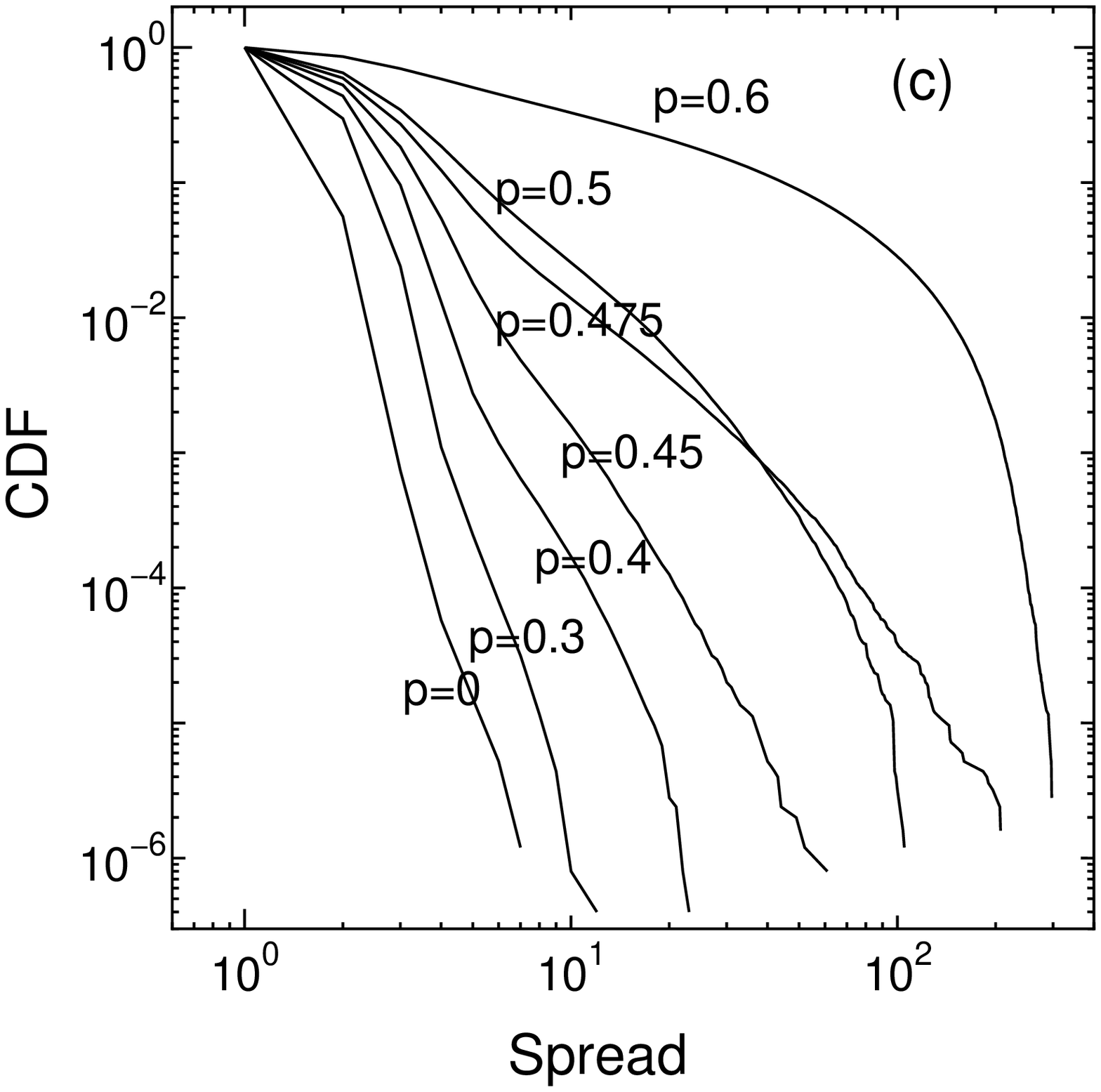}
\caption{Cumulative distribution functions of price shifts, the
gaps between ask and the second best sell limit price and spreads.
The power law exponents of price shifts (the gaps, spreads) are
$3.97\pm0.11$ ($4.27\pm0.12$, $4.49\pm0.11$), $2.72\pm0.08$
($2.97\pm0.11$, $3.09\pm0.08$) and $3.78\pm0.11$ ($3.80\pm0.11$,
$4.14\pm0.10$) for $p$=0.45, 0.475 and 0.5 respectively.}
\label{fig1}
\end{figure*}

\newpage

\begin{figure}
\includegraphics[width=8cm]{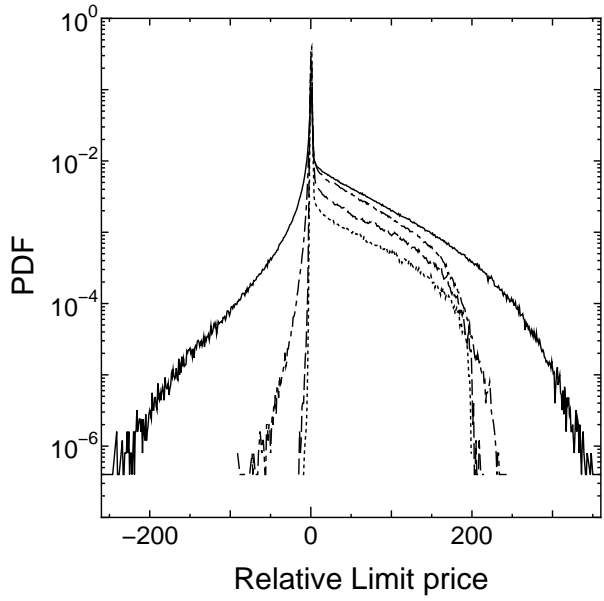}
\caption{Probability distribution function of the relative limit
price. The results are shown for the three cases with $p$=0.3
(dotted line), $p$=0.4 (dashed line), $p$=0.5 (dot-dash line) and
$p$=0.6 (solid line).} \label{fig2}
\end{figure}

\begin{figure}
\includegraphics[width=8cm]{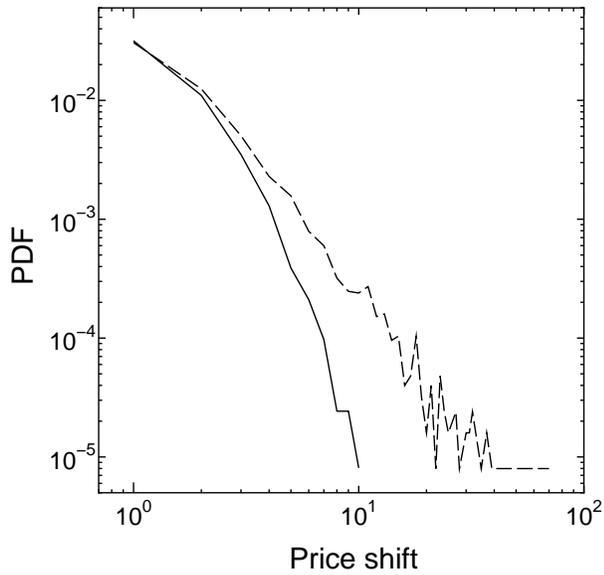}
\caption{Probability distribution function of price shift of the
surrogate data. The original data is generated by 1,00 times runs
of 10,000 step iterations with $p$=0.5. The comparison with that
of the original data (dashed line) is also given.} \label{fig3}
\end{figure}

\newpage

\begin{figure*}
\includegraphics[width=8cm]{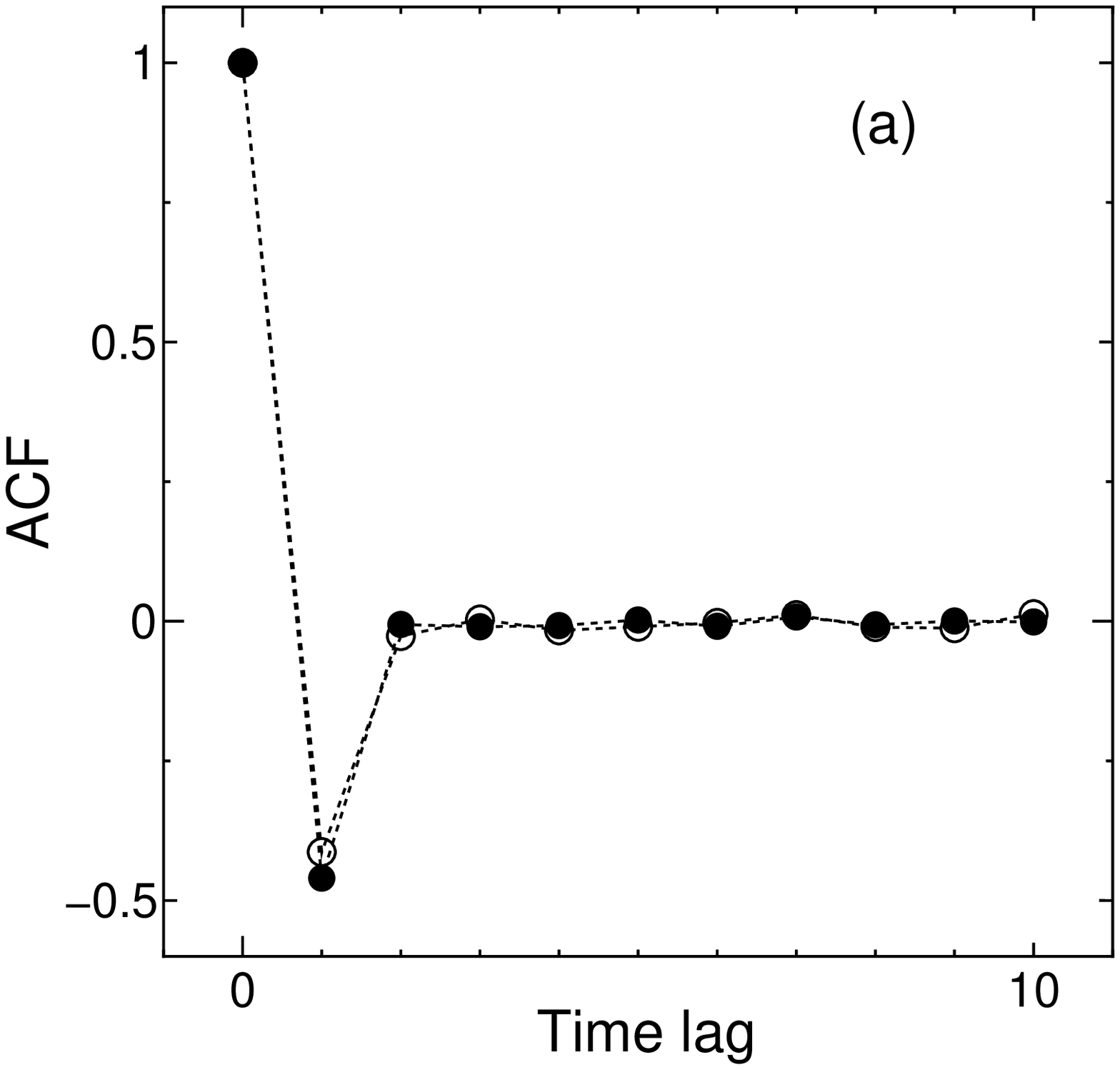}
\end{figure*}
\begin{figure*}
\includegraphics[width=8cm]{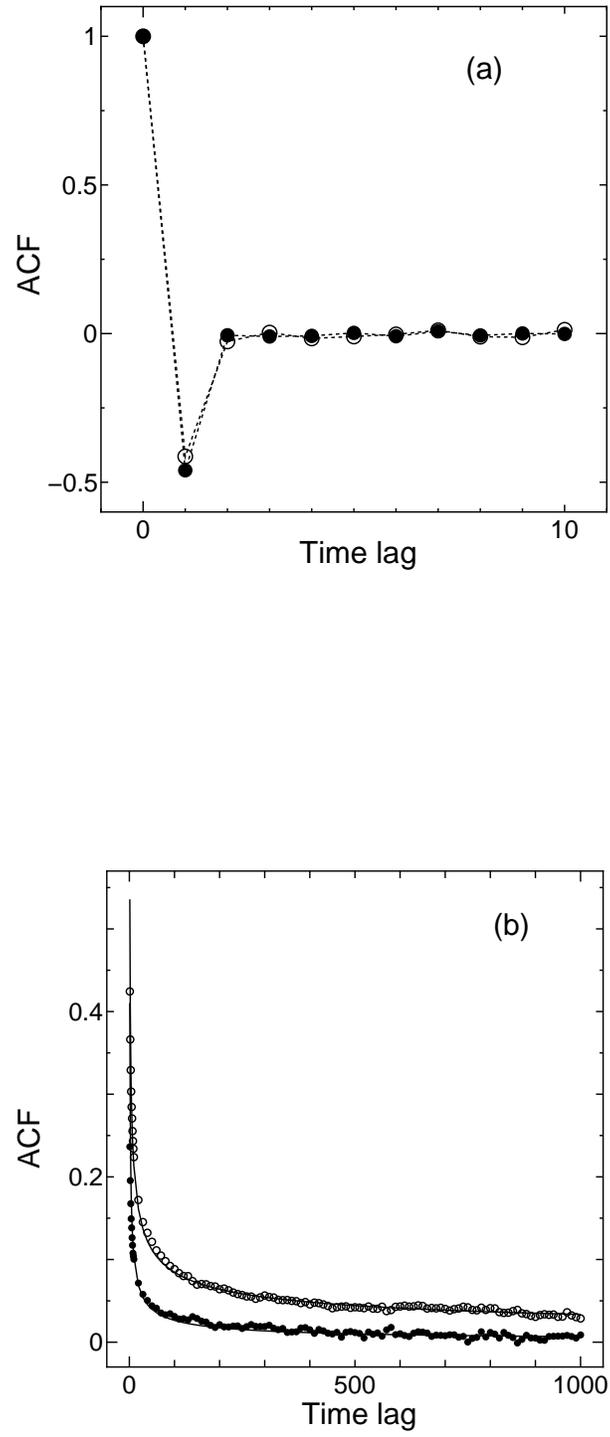}
\caption{Autocorrelation functions of price shift and of the
absolute value of price shift obtained by the numerical
simulations of the model. In both panels, the unit of time
increment corresponds to a buy market order. Empty circle
($\circ$) represents the results for the original data, and filled
circle ($\bullet$) for the surrogate data mentioned in the
previous section. (a)The autocorrelation function of price shift.
(b)The autocorrelation of the absolute value of price shift with
the power law fittings(solid lines). The exponents of the power
law fittings are estimated by linear regression of the data
plotted in log-log plain. The result is -0.40 ($R^2=0.99$) for the
original data, and -0.60 ($R^2=0.79$) for the surrogate data.}
\label{fig4}
\end{figure*}

\newpage

\begin{figure}
\includegraphics[width=8cm]{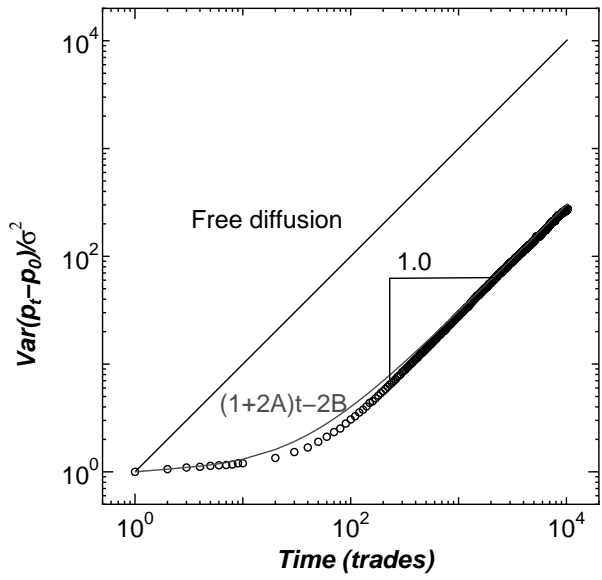}
\caption{Empirical study of the price diffusion. We analyzed about
45 millions transaction data from Nov. 1999 through Oct. 2000 of
active 5 IT or e-commerce companies (Intel, Microsoft, Amazon,
Oracle, Cisco) listed on Nasdaq using TAQ Database. The
theoretical line is also given, where $A=\sum_{i=1}^t\rho_i$ and
$B=\sum_{i=1}^t i\rho_i$.} \label{fig5}
\end{figure}

\end{document}